\documentclass[twocolumn,pra,aps,showpacs,amsmath,amssymb]{revtex4-1}
\usepackage{physics}
\usepackage{amsmath}
\usepackage{mathrsfs}
\usepackage{bbold}
\usepackage{graphicx}
\usepackage{color}
\usepackage{bm}
\usepackage{soul}
\usepackage{ulem}
\usepackage{bbm}
\usepackage{bigints}
\newcommand{\s}[1]{\scriptscriptstyle{#1}}

\graphicspath{{NewFigures/}}

\usepackage{bigints}

\begin{document}


\title{Space-time-symmetric extension of quantum mechanics: Interpretation and arrival-time predictions}

\author{Ruben E. Araújo$^1$}
 
\author{Ricardo Ximenes$^2$}%
\author{Eduardo O. Dias$^1$}
\email{eduardo.dias@ufpe.br}
\affiliation{%
		$^1$Departamento de F\'{\i}sica, Universidade Federal de Pernambuco,
		Recife, Pernambuco  50670-901, Brazil}
	\affiliation{%
		$^2$Department of Physics, University of Wisconsin-Madison, Madison, Wisconsin 53706, USA.}
\date{\today}

\begin{abstract}
An alternative quantization rule, in which time becomes a self-adjoint operator and position is a parameter, was proposed by Dias and Parisio [Phys. Rev. A {\bf 95}, 032133 (2017)]. In this approach, the authors derive a space-time-symmetric (STS) extension of quantum mechanics (QM) where a new quantum state (intrinsic to the particle), $|{\phi}(x)\rangle$, is defined at each point in space. $|\phi(x)\rangle$ obeys a space-conditional (SC) Schr\"odinger equation and its projection on $|t\rangle$, $\langle t|\phi(x)\rangle$, represents the probability amplitude of the particle's arrival time at $x$. In this work, first we provide an interpretation of the SC Schr\"odinger equation and the eigenstates of observables in the STS extension. Analogous to the usual QM, we propose that by knowing the ``initial'' state  $|\phi(x_0)\rangle$ --- which predicts any measurement on the particle performed by a detector localized at $x_0$ --- the SC Schr\"odinger equation provides $\ket{\phi(x)}=\hat{\mathbbm U}(x,x_0)\ket{\phi(x_0)}$, enabling us to predict measurements when the detector is at $x \lessgtr x_0$. We also verify that for space-dependent potentials, momentum eigenstates in the STS extension, $|P_b(x)\rangle$, depend on position just as energy eigenstates in the usual QM depend on time for time-dependent potentials. In this context, whereas a particle in the momentum eigenstate in the standard QM, $|\psi(t)\rangle=|P\rangle|_t$, at time $t$, has momentum $P$ (and indefinite position), the same particle in the state $|\phi(x)\rangle=|P_b(x)\rangle$ arrives at position $x$ with momentum $P_b(x)$ (and indefinite arrival time). By investigating the fact that $|\psi(t)\rangle$ and $|{\phi}(x)\rangle$ describe experimental data of the same observables collected at $t$ and $x$, respectively, we conclude that they provide complementary information about the same particle. Finally, we solve the SC Schr\"odinger equation for an arbitrary space-dependent potential. We apply this solution to a potential barrier and compare it with a generalized Kijowski distribution, showing that they can predict distinct traversal times.
\end{abstract}



\pacs{Valid PACS appear here}

\maketitle

\section{\label{sec:level1}Introduction}

In ordinary quantum mechanics (QM), time is a parameter, $t$, and position is a self-adjoint operator, ${\hat {\rm X}}$. In this asymmetric formulation, the state of a particle represented in the eigenstates of ${\hat {\rm X}}$ provides the probability amplitude, $\psi(x|t)\equiv \langle x| \psi(t)\rangle$, of finding this particle at position $x$, {\it given that} the measurement takes place at time $t$. Because of this time-conditional (TC) character of QM, it will be convenient for us to use the notation $(x|t)$. This asymmetry is in part responsible for serious conflicts in the foundations of QM, including the problem of time in attempts to unify QM and general relativity~\cite{Zeh,Anderson}, the interpretation of the time-energy uncertainty relation~\cite{Time}, and the predictions of arrival and tunneling times~\cite{Time,Time2,Muga}.

Aiming for a more symmetrical treatment for space and time in QM, Dias and Parisio in Ref.~\cite{Dias} proposed a complementary way of quantizing classical variables. Position becomes a parameter and time a self-adjoint operator $\mathbbm {\hat T}$ obeying the canonical commutation relation $[{\hat {\mathbbm H}},{\hat {\mathbbm T}}]=i\hbar$. Here ${\hat {\mathbbm H}}$ is a Hamiltonian operator different from the usual one, even though both come from the same classical Hamiltonian. In addition to the TC state $|\psi(t)\rangle$, Ref.~\cite{Dias} postulated the existence of a space-conditional (SC) state, $|{\phi}(x)\rangle$, intrinsic to the particle and defined at every position in space. Its wave function ${\pmb \phi}(t|x)$ is a two-component vector that  represents [analogously to $\psi(x|t)$ under $x \rightleftarrows t$] the probability amplitude of measuring the particle at time $t$, given that the observation happens at position $x$.

From this space-time-symmetric (STS) extension of QM, Ref.~\cite{Dias} proposed how $|{\phi}(x)\rangle$ changes in space, similarly to how $|\psi(t)\rangle$ evolves in time. From now on, we will call this ``dynamic'' constrain the space-conditional (SC) Schr\"odinger equation, which was also studied in Refs.~\cite{Ricardo,Lara,Dias3,Lara2}. For the free-particle case~\cite{Dias}, this equation resulted (with a particular assumption that will be discussed later) in Kijowski's time of arrival (TOA) distribution~\cite{Kijo}, which has been derived via different methods using the usual QM~\cite{All,Aharonov,Grot,Delgado}.

The ordinary TOA problem consists of predicting the time probability distribution for a particle to arrive at a given position $x>0$, given its initial wave packet, $\psi(x|t_0)$, restricted to $x<0$. The existing models of this problem can be divided into classes: independent~\cite{Dias,Kijo,Aharonov,Grot,Galapon,Galapon1,Galapon2,Delgado,Dumont,Leavens2,Leavens3,Daumer,Leavens4,Grubl,Das,Das2,Vona1,Das3,Vona,Anantha,Leavens,Kazemi,Kazemi2}  and dependent on the measurement device~\cite{All,MugaComplex,Halliwell,MugaComplex2,Echanobe,Nikolic,Jad,Jad2,Schuss,Maccone,Werner,Tumulka,Maccone2,Dias2}. The former distributions are an intrinsic property of the particle state and aim to describe the so-called ideal TOA.

A common approach to the ideal TOA~\cite{Aharonov,Grot,Galapon,Galapon1,Galapon2,Delgado} assumes that, like any observable in QM, it can be determined by the spectral decomposition of a time operator and the measurement-free state of the particle state, $|\psi(t)\rangle$. Nevertheless, such a procedure faces the Pauli objection~\cite{Pauli}: a self-adjoint time operator, ${\hat {\rm T}}$, satisfying $[{\hat {\rm H}},{\hat {\rm T}}]=i\hbar$ requires a Hamiltonian unbounded from below, which is unphysical in the usual QM. As a result, these ideal models are compelled to abdicate either the canonical commutation relation with the Hamiltonian or the self-adjointness of the time operator~\cite{Delgado}. Because of this objection, the theoretical description of arrival, traversal, and tunneling times remain controversial, with multiple approaches emerging in recent decades. As it will be clear later, the Pauli argument does not apply to the STS extension because of its distinct quantization rules.

Another traditional model of the ideal TOA is the quantum flux (probability current) density~\cite{Dumont,Leavens2,Leavens3,MugaComplex,Daumer,Leavens4,Grubl,Das,Das2,Vona1,Das3,Vona,Anantha,Leavens}, which, like other approaches, is not unrestrictedly valid~\cite{Kijo,Vona1,Leavens,Leavens2,Das2,Vona}. For instance, it is known that the quantum flux can predict negative probabilities, an effect called quantum backflow~\cite{Bracken,Bracken2,Halli}, and cannot always be described by a POVM (positive-operator valued measure).

The numerous disparate models of the TOA in the literature, as well as the appealing symmetrical formulation of the STS extension, motivated us to develop this theory. Besides that, the STS extension has been shown to be a promising approach to the arrival and tunneling time problems~\cite{Dias,Ricardo,Dias3,Lara}. In particular, Ref.~\cite{Dias3} generalizes the STS extension to a particle moving in three-dimensional space, and discusses how it can be extended to the relativistic domain. The results of Ref.~\cite{Ricardo} will be mentioned later.


In this work, we propose an interpretation of the STS extension and we solve the SC Schr\"odinger equation for an arbitrary time-independent potential. This solution is applied to a potential barrier. This manuscript is organized as follows. In Sec.~\ref{review}, we review in detail the STS extension, and in Sec.~\ref{comparison}, we investigate how to represent the solutions, $|\phi(x)\rangle$, of the SC Schr\"odinger equation for arbitrary potentials in the momentum basis, which plays an equivalent role to the energy basis in the usual QM. In Sec.~\ref{interpretation}, we propose an interpretation of these solutions and the eigenstates (along with their wave functions) of any observable in the STS extension. Finally, in Sec.~\ref{solution}, we solve the SC Schr\"odinger equation for an arbitrary time-independent potential. We apply this solution to a potential barrier and compare it with a generalized Kijowski distribution. As our focus is on the interpretation of the STS theory, comparisons between our solution with other time-of-arrival distributions and the experiments of Ref.~\cite{Steinberg} are left for future work.

\section{Review of the STS extension of QM}
\label{review}

Let us review the STS extension of QM by using a notation similar to that of Ref.~\cite{Dias} and drawing a parallel with some basic concepts of QM. It is worth mentioning that the original work on the STS extension, Ref.~\cite{Dias}, is a relatively short article, so a more detailed formulation of this theory is still lacking.

The goal of the STS extension proposed in Ref.~\cite{Dias} is to deal with experimental situations complementary to those of the usual QM, which involve TC distributions {\it intrinsic to the particle}, $|\psi(x|t)|^2$. For instance, one could ask about the joint probability ${\cal P}(x,t)dxdt$ of finding the particle in a given region of space $[x,x+dx]$ and in a certain interval of time $[t,t+dt]$. In this situation, ${\cal P}(x,t)$ is equal to the probability density of finding the particle at position $x$ given that the observation occurs precisely at $t$, ${\cal P}(x|t)=|\psi(x|t)|^2$, times the probability density ${\cal P}(t)$ of the system being measured at instant $t$, whatever the outcome. Thus, we have
\begin{eqnarray}
\label{P1}
{\cal P}(x,t)dxdt={\cal P}(x|t){\cal P}(t)dxdt=|\psi(x|t)|^2{\cal P}(t)dxdt,
\end{eqnarray}
where ${\cal P}(x,t)$ and ${\cal P}(t)$ cannot be obtained exclusively through $|\psi(x|t)|^2$. It is worth remarking that the last equality of Eq.~(\ref{P1}) assigns to ${\cal P}(x|t)$ the modulus square of a complex function. This relationship, together with the linearity of QM, differentiates QM from classical probability theories. These features allow probabilities such as $|\psi_1(x|t)+\psi_2(x|t)|^2$, yielding the interference phenomenon in the possible positions where one can find the particle. From Eq.~(\ref{P1}), Ref.~\cite{Dias} defines a global wave function $\Psi(x,t)$ whose modulus square is the joint probability distribution of finding the particle at $x$ and $t$, ${\cal P}(x,t)=|\Psi(x,t)|^2$, and is normalizable by integration over space and time.

The STS extension emerges by using Bayes' theorem, which allows ${\cal P}(x,t)$ in Eq.~(\ref{P1}) to be rewritten as
\begin{eqnarray}
\label{P2}
{\cal P}(x,t)dxdt={\cal P}(t|x){\cal P}(x)dxdt,
\end{eqnarray}
where ${\cal P}(t|x)$ is the probability density of finding the particle at $t$, given that the observation takes place at position $x$, and ${\cal P}(x)$ is the probability distribution of finding the particle at $x$, irrespective of the time the observation happens. In Ref.~\cite{Dias}, ${\cal P}(t|x)$ is interpreted as the TOA distribution at position $x$. Note that in Eq.~(\ref{P1}), $x$ and $t$ play opposite roles in comparison with Eq.~(\ref{P2}). In this context, the STS extension conjectures that the time probability distribution ${\cal P}(t|x)$, analogously to the space distribution $P(x|t)=|\psi(x|t)|^2$, comes from the modulus square of a complex function {\it intrinsic to the particle}, but now conditioned at position $x$:
\begin{eqnarray}
\label{P22}
{\cal P}(t|x) \equiv N|{\pmb \phi}(t|x)|^2,
\end{eqnarray}
where $N$ is a normalization factor and the use of boldface will be clear later. In this manner, the interference phenomenon for the TOA of the particle emerges naturally. From this perspective, the usual QM can be seen as a {\it time-conditional} (TC) QM and its STS extension as a {\it space-conditional} (SC) QM. This work will focus on the SC wave function ${\pmb \phi}(t|x)$ rather than the global wave function $\Psi(x,t)$.

With this initial discussion, we have the physical intuition to define the mathematical elements of the STS extension. First, to make a parallel with QM, we review some of its basic properties. In the usual QM, the state of a one-dimensional spinless particle, $|\psi(t)\rangle$, is defined at every instant $t$ and belongs to a Hilbert space ${\cal H}|_t$. The notation $|_t$ does not mean a time dependence; instead, it emphasizes that time is a parameter in this space, i.e., its physical states, $|\psi(t)\rangle$, are conditioned on a given time. Here, position is an operator acting on ${\cal H}|_{t}$ such that
\begin{equation}\label{position}
{\hat {\rm X}} |x\rangle=x|x\rangle~~{\rm and}~~ [{\hat {\rm X}},{\hat {\rm P}}]=i\hbar,
\end{equation}
where $\langle x|x' \rangle=\delta (x-x')$ and ${\hat {\rm P}}$ is the momentum operator. The commutation relation~(\ref{position}) leads to the uncertainty principle of position and momentum, $\Delta {\hat {\rm X}} \Delta {\hat {\rm P}} \geq \hbar/2 $, 
where $\Delta$ is the root-mean-square deviation. In the position representation, $\hat {\rm P}$ is
\begin{equation}\label{momentum}
 \langle  x|{\hat {\rm P}}| x' \rangle=-i\hbar \frac{\partial}{\partial x}\delta(x-x')
\end{equation}
and its eigenstate $|P\rangle \equiv |P\rangle|_t$, with ${\hat {\rm P}}|P\rangle|_t= P |P\rangle|_t$, is
\begin{equation}\label{autoestadoP}
|P\rangle|_t=\frac{1}{\sqrt{2\pi \hbar}} \int_{-\infty}^{\infty} dx~ e^{iPx/\hbar} |x\rangle.
\end{equation}
As it will be clear in Sec.~\ref{interpretation}, the notation $|\rangle|_t$ will be essential to distinguish between eigenstates of the same classical observable in the usual QM and the STS extension ($|\rangle|_x$), which have different physical meanings. Again, $|\rangle|_t$ does not mean a time dependence. Here the resolution of the identity reads
\begin{equation}\label{identityX}
\int_{-\infty}^{\infty} dx ~|x\rangle \langle x|=\int_{-\infty}^{\infty} dP~ |P\rangle|_t|\langle P|= \mathbbm{1}.
\end{equation}

The physical information about the particle position at instant $t$ is contained in $|\psi(t)\rangle$ via the expansion
\begin{equation}\label{expansionX}
|\psi(t)\rangle=\int_{-\infty}^{\infty} dx~ \psi(x|t) |x\rangle,
\end{equation}
which is the solution of the Schr\"odinger equation,
\begin{equation}\label{Schro}
{\hat {\rm H}}|\psi(t)\rangle=i\hbar \frac{d}{dt}|\psi(t)\rangle.
\end{equation}
Here ${\hat {\rm H}}$ is obtained from the classical Hamiltonian via the quantization rule $(x,P) \rightarrow ({\hat {\rm X}},{\hat {\rm P}})$, i.e.,
\begin{equation}\label{ruleX}
H(x,P;t)=\frac{P^2}{2m}+V(x,t) ~ \rightarrow ~ {\hat {\rm H}}({\hat {\rm X}},{\hat {\rm P}};t)=\frac{{\hat {\rm P}}^2}{2m}+{\hat {\rm V}}({\hat {\rm X}},t).
\end{equation}
Substituting Eq.~(\ref{ruleX}) into Eq.~\eqref{Schro}, and projecting the resulting expression on $|x\rangle$, we obtain
\begin{equation}\label{Schro2}
\left[-\frac{\hbar^2}{2m}\pdv[2]{x}+V(x,t)\right]\psi(x|t)=i\hbar \pdv{\psi (x|t)}{t}.
\end{equation}
Finally, the probability of finding the particle in the region $[x,x+dx]$ given that the measurement takes place at time $t$ is
\begin{equation}\label{rhoX}
{\cal P}(x|t) dx=|\langle x|\psi(t)\rangle|^2dx=\psi^{*}(x|t)\psi(x|t) dx.
\end{equation}

Now, we turn our attention to the STS extension proposed in Ref.~\cite{Dias}. For the sake of understanding, let us formulate it following the same steps as above for QM. In the STS extension, a distinct quantization rule is performed. The state of a one-dimensional spinless particle, $|\phi(x)\rangle$, is defined at each position $x$ and belongs to a Hilbert space ${\cal H}|_x$. Like $|_t$ in the usual QM, $|_x$ does not mean a space dependence but that the physical states, $|\phi(x)\rangle$, belonging to ${\cal H}|_x$ are conditioned on a given position. Time is an operator, 
${\hat {\mathbbm T}}$, acting on ${\cal H}|_x$ and canonically conjugated to the Hamiltonian operator ${\hat {\mathbbm H}}$ ($\neq {\hat {\rm H}}$), i.e.,
\begin{equation}\label{time}
{\hat {\mathbbm T}}|t\rangle =t|t\rangle~~{\rm and}~~ [{\hat {\mathbbm H}}, {\hat {\mathbbm T}}]=i\hbar,
\end{equation}
where $\langle t|t' \rangle=\delta (t-t')$. This commutation relation leads to the time-energy uncertainty relation,
 $\Delta {\hat {\mathbbm T}} \Delta {\hat {\mathbbm H}} \geq \hbar/2 $.

It is important not to confuse ${\hat {\mathbbm H}}$ acting on ${\cal H}|_x$ with ${\hat {\rm H}}$ (Eq.~(\ref{ruleX})) acting on ${\cal H}|_t$, although both refer to the same Hamiltonian of classical mechanics. They arise from distinct quantization rules and act upon different Hilbert spaces. In the time representation, ${\hat {\mathbbm H}}$ is defined as
\begin{equation}\label{energy}
\langle t|{\hat {\mathbbm H}}|t'\rangle  =i\hbar \frac{\partial}{\partial t}\delta(t-t')
\end{equation}
and its eigenstate $|E\rangle|_x$ (${\hat {\mathbbm H}}|E\rangle|_x= E |E\rangle|_x$) becomes
\begin{equation}\label{autoestadoH}
|E\rangle|_x=\frac{1}{\sqrt{2\pi \hbar}} \int_{-\infty}^{\infty}dt~ e^{-iE t/\hbar} |t\rangle.
\end{equation}
In this Hilbert space, the resolution of the identity is
\begin{equation}\label{identityT}
\int_{-\infty}^{\infty} dt ~|t\rangle \langle t|=\int_{-\infty}^{\infty} dE~ |E\rangle|_x|\langle E|= \mathbbm{1}.
\end{equation}
Comparing Eqs.~(\ref{position})-(\ref{identityX}) with Eqs.~(\ref{time})-(\ref{identityT}), we observe that, as with space and time, energy (momentum) in ${\cal H}|_t$ plays the same role as momentum (energy) in ${\cal H}|_x$. From Eq.~(\ref{identityT}), $E$ can be {\it a priori} negative, in the same way as $P$ in Eq.~(\ref{identityX}). In this scenario, just as the amplitudes of $|\psi(t)\rangle$ in the basis $\{|P\rangle|_t\}$, ${_t |}\langle P|\psi(t)\rangle$, select the possible momenta of the system (depending on the potential, and initial and boundary conditions), the amplitudes of $|{\phi}(x)\rangle$ in the basis $\{|E\rangle|_x\}$, ${_x |}\langle E|\phi(x)\rangle$, select the energies of the system. Since in the very formulation of the theory, the basis $\{|E\rangle|_x\}$ includes energies from minus to plus infinity, the Pauli objection does not apply to the Hilbert space of the STS extension.

Analogous to Eq.~(\ref{expansionX}), with $x \rightleftarrows t$, the physical information about the TOA of the particle at position $x$ is contained in $|{\phi}(x)\rangle$ via the expansion
\begin{equation}\label{expansionT1}
\ket{{\phi}(x)}=\int_{-\infty}^{\infty} dt~ {\pmb \phi}(t|x) ~|t\rangle.
\end{equation}
The physical constraint of $\ket{{\phi}(x)}$, defined below, makes the SC wave function, ${\pmb \phi}(t|x)$, a two-component object,
\begin{equation}\label{solT}
{\pmb \phi}(t|x)=\mqty(\phi^{+}(t|x)  \\  \phi^{-}(t|x)).
\end{equation}
By defining the kets $ |+\rangle \equiv \scriptscriptstyle{  \mqty(1\\0)} $ and $ |-\rangle \equiv \s{  \mqty(0\\1)} $, Eq.~(\ref{expansionT1}) can be written as
\begin{equation}\label{expansionT2}
\ket{{\phi}(x)}=\int_{-\infty}^{\infty} dt~[ {\phi}^+(t|x) ~|t\rangle\otimes |+\rangle +{\phi}^-(t|x) ~|t\rangle\otimes |-\rangle].
\end{equation}

Since one takes $x \rightleftarrows t$ (and $P \rightleftarrows E$) in QM to formulate the STS extension, $|{ \phi}(x)\rangle$ should change in space analogously to how $|\psi(t)\rangle$ evolves in time via the Schr\"odinger equation~(\ref{Schro}). As described above, in the usual QM, the generator of time translations, ${\hat {\rm H}}$, is obtained by quantizing the classical Hamiltonian, Eq.~(\ref{ruleX}). To obtain the corresponding generator of space translations in the STS extension, we should apply the new quantization rules [Eqs.~(\ref{time}) and~(\ref{energy})] to the classical momentum, i.e.,
\begin{eqnarray}\label{ruleP}
P(t,H;x)&=&\pm \sqrt{2m\big [H-V(x,t)\big ]} \nonumber\\
 \rightarrow   {\hat {\mathbbm P}}({\hat {\mathbbm T}},{\hat {\mathbbm H}};x)&=&\sigma_{z} \sqrt{2m\left[{\hat {\mathbbm H}}-V(x,{\hat {\mathbbm T}})\right]},
\end{eqnarray}
where $\sigma_z={\rm diag}(+1,-1)$. Then, for the space translation of $|{\phi}(x)\rangle$ to be analogous to the time translation of $|\psi(t)\rangle$, $|{ \phi}(x)\rangle$ should obey
\begin{equation}\label{SchroT}
{\hat {\mathbbm P}}\ket{{\phi}(x)}=-i\hbar \frac{d}{dx}\ket{{\phi}(x)},
\end{equation}
which we call the SC Schr\"odinger equation. In the time representation, $\{|t\rangle\}$, Eq.~(\ref{SchroT}) reads
\begin{equation}\label{Schro2T}
\sigma_{z} \sqrt{2m\left(i\hbar \pdv{t}-V(x,t)\right)}{\pmb \phi}(t|x)=-i\hbar \frac{\partial {\pmb \phi}(t|x)}{\partial x}.
\end{equation}
with ${\pmb \phi}(t|x)$ given by Eq.~(\ref{solT}). Analogous to $t$ in the usual QM, $x$ in the STS extension is a continuous parameter that one can choose with arbitrary precision to evaluate the time probability amplitude ${\pmb \phi}(t|x)$. Similar to what happens with ${\hat {\rm H}}$ and $t$, ${\hat {\mathbbm P}}$ and $x$ cannot satisfy the standard uncertainty principle.

From ${\pmb \phi}(t|x)$, we obtain the probability for the particle to arrive at $x$ in $[t,t+dt]$ via
\begin{equation}\label{rhoT}
{\cal P}(t|x)dt=\frac{|\langle t|{ \phi}(x)\rangle|^2}{\langle {\phi}(x)|{ \phi}(x) \rangle} dt=\frac{{\pmb \phi}^{\dagger}(t|x){\pmb \phi}(t|x)}{\langle {\phi}(x)|{\phi}(x) \rangle}dt,
\end{equation}
where the symbol $\dagger$ is the conjugate transpose operation. In Eq.~(\ref{rhoT}), the normalization with ${\langle {\phi}(x)|{ \phi}(x) \rangle}$ is necessary because ${\hat {\mathbbm P}}$ is not always Hermitian. As a result, the SC Schr\"odinger equation~(\ref{SchroT}) is not unitary in general. Note that ${\langle { \phi}(x)|{ \phi}(x) \rangle}$ is the probability for the particle to arrive at $x$ regardless of the TOA. Although we can observe a particle at any instant of time (admitting that it exists, $\langle \psi(t)|\psi(t)\rangle=1$), we cannot observe it at any position, even if we wait for an infinite time interval. Therefore, we can have $0 \leq \langle { \phi}(x)|{\phi}(x) \rangle \leq 1$~\cite{Ricardo}. Here, an essential difference between ${\cal H}|_t$ and ${\cal H}|_x$ becomes evident: The physical states in ${\cal H}|_t$ (${\cal H}|_x$) have square-integrable wave functions in space (time).

The formulation above of the STS extension does not involve the quantum states of detectors and/or clocks that measure the TOA, but only the properties of the particle itself. In this manner, ${\pmb \phi}(t|x)$ can be identified as a probability amplitude of an ideal TOA. At this point, it is worth noting the difference between the well-known Page and Wootters formalism~\cite{Page,Seth,Vedral,Boette} and the STS extension. The former considers an extra physical system playing the role of a clock, with a time superposition referring to the history of the system. In contrast, in the STS extension, the time superposition refers to a single event, the particle's TOA.

Reference~\cite{Dias} solves Eq.~(\ref{Schro2T}) only for the free particle situation, $V(x,t)=0$. Identifying $\sqrt{d/dt}$ with the Riemann-Liouville fractional derivative $_{-\infty}D^{1/2}|_t$, which is equivalent to the Caputo fractional derivative~\cite{frac}, we have $_{-\infty}D^{1/2}_t \exp(-iwt)=\sqrt{-iw} \exp(-iwt)$. Under these circumstances, the time probability density (\ref{rhoT}) becomes
\begin{eqnarray} \label{pd}
{\cal P}(t|x)= \frac{1}{2\pi m\hbar}
\bigg\{{\bigg|}\int_0^{\infty}{\tilde \phi}(+P)\sqrt{P}{\rm
e}^{iPx/\hbar-iE_P t/\hbar}dP{\bigg |}^2
\nonumber\\
+ {\bigg |} \int_0^{\infty}{\tilde \phi}(-P)\sqrt{P}{\rm e}^{-iPx/\hbar -
iE_P t/\hbar}dP{\bigg |}^2 ~\bigg\}\frac{1}{\langle \phi(x)|\phi(x) \rangle},\nonumber\\
\end{eqnarray}
where $|{\tilde \phi}(P)|^2$ is the probability density of the particle having momentum $P $ (where $P$ is a real number), given that its observation happens at position $x$. Reference~\cite{Dias} recognizes Eq.~(\ref{pd}) as the normalized Kijowski TOA distribution~\cite{Kijo} by identifying ${\tilde \phi}( P)$ as the momentum wave function in the usual QM, ${\tilde \psi}(P)$. Nevertheless, since probabilities in the STS extension are conditioned on a given position, this identification requires a more careful investigation. Later, after giving an interpretation of the STS extension, we will discuss the consequences of assuming ${\tilde \phi}(P)={\tilde \psi}(P)$.


\section{ $|\psi(t)\rangle$ in the energy basis  {\it versus} $|\phi(x)\rangle$ in the momentum basis}\label{comparison}

From Eqs.~(\ref{Schro}) and~(\ref{SchroT}), we verify that the energy eigenvalue equation in the usual QM plays an equivalent role to the momentum eigenvalue equation in the STS extension. Thus, to help understand the interpretation proposed in the next section, let us compare $|\psi(t)\rangle$ and $|{ \phi}(x)\rangle$ in the energy and momentum bases, respectively, for an arbitrary potential $V=V(x,t)$. It is worth emphasizing that Ref.~\cite{Dias} focused solely on the free-particle case, whereas the discussion in this section concerns $V \neq 0$.

In the usual QM, the instantaneous energy eigenstates satisfy~\cite{Sakurai}
\begin{equation}\label{SchroE1}
{\hat {\rm H}}(t)|E_a(t)\rangle = E_a(t)|E_a(t)\rangle,
\end{equation}
where $\langle E_{a'}(t)|E_a(t)\rangle=\delta_{a'a}$ and $\sum_a |E_a(t)\rangle{\langle} E_a(t)|=1$. Similar to $|\psi(t)\rangle$, we are omitting  $|_t$ in $|E_a(t)\rangle$. An arbitrary state in ${\cal H}|_t$ can then be expanded in this instantaneous energy basis,
\begin{equation}\label{expansion}
    \ket{\psi(t)}=\sum_a ~ {\bar \psi}(a|t) \ket{E_a(t)},
\end{equation}
where
\begin{equation}\label{PsiE}
    {\bar \psi}(a|t) \equiv {\langle} E_a(t)|\psi(t)\rangle= C(a|t)e^{i \theta_a(t)},
\end{equation}
with $\theta_a(t)=-1/\hbar\int_0^t \dd{t'}E_a(t')$
being the dynamical phase.  From the Schr\"odinger equation~(\ref{Schro}), it is straightforward to see that the coefficient $C(a|t)$ satisfies
\begin{equation}\label{Ca}
    \frac{dC(a|t)}{dt}=-\sum_{a'} C(a'|t)e^{i\Delta \theta_{a'a}(t)}\bra{E_a(t)}\frac{d}{dt}\ket{E_{a'}(t)},
\end{equation}
where we introduced $\Delta \theta_{a'a}(t)=\theta_{a'}(t)-\theta_a(t)$. Note that ${\cal P}(a|t)=|{\bar \psi}(a|t)|^2=|C(a|t)|^2$ ($\theta_{a}(t)$ is a real number) is the probability density of measuring the particle in the state $|E_a(t)\rangle$ (with energy $E_{a}(t)$), given that the measurement happens at time $t$. Now, projecting Eq.~(\ref{expansion}) on $|x\rangle$ yields
\begin{equation}\label{SolX3}
\psi(x|t)=\langle x|\psi(t)\rangle=\sum_a ~{\bar \psi}(a|t) \psi_a(x|t),
\end{equation}
where  $\psi_a(x|t)=\langle x|E_a(t)\rangle$ is the probability amplitude for the particle to be found at position $x$, given that the observation happens at time $t$ and its state is $|E_a(t)\rangle$. For the Hamiltonian~(\ref{ruleX}), the projection of Eq.~(\ref{SchroE1}) on $ \ket{x}$ gives
\begin{equation}\label{SchroXE}
    \left[-\frac{\hbar^2}{2m}\pdv[2]{x}+V(x,t) \right]\psi_a(x|t)=E_a(t) ~\psi_a(x|t).
\end{equation}

In particular, for the free particle situation, where $[{\hat{\rm H}},\hat{\rm P}]=0$, the energy eigenstate~(\ref{SchroE1}) becomes time-independent: $|E_a(t)\rangle=|E;\pm\rangle|_t$,  where $\pm$ refers to the sign of the momentum and $E={P}^2/(2m)$. Also, ${\bar \psi}(a|t)$ of Eq.~(\ref{PsiE}) can be written as ${\bar \psi}(E;\pm|t)={\bar \psi}(E;\pm) \exp\{-iEt/\hbar\}$. Thus, Eq.~(\ref{expansion}) can be rewritten as
\begin{eqnarray}\label{SolXL}
&&|\psi(t)\rangle\nonumber=\\
&&\int_{0}^{\infty} dE  \big[{\bar \psi}(E;+) |E;+\rangle|_t+{\bar \psi}(E;-)  |E;-\rangle|_t \big]e^{-iEt/\hbar},\nonumber\\
\end{eqnarray}
with ${\cal P}(E;\pm|t)=|{\bar \psi}(E;\pm)|^2$. Finally, the energy eigenfunction
 $\psi_a(x|t)=\langle x|E_a(t)\rangle|_t$ also becomes time independent,
\begin{eqnarray}\label{SolXFree}
\psi_{E;\pm}(x)=\langle x|E;\pm\rangle|_t=\frac{1}{\sqrt{2\pi\hbar}}\left(\frac{m}{2 E  }\right)^{1/4}~e^{\pm i\sqrt{2mE}x/\hbar}.\nonumber\\
\end{eqnarray}

The equivalent analysis for $V=V(x,t)$ in the STS extension starts with the momentum eigenvalue equation,
\begin{equation}\label{SchroTP1}
{\hat {\mathbbm P}(x)}\ket{P_b(x)}= P_b(x)\ket{P_b(x)},
\end{equation}
where
\begin{eqnarray}\label{Pb}
\ket{P_b(x)}=|P_b^+(x)\rangle \otimes |+\rangle +|P^-_b(x)\rangle \otimes |-\rangle= \mqty(|P_b^+(x)\rangle \\ |P^-_b(x)\rangle),\nonumber\\
\end{eqnarray}
with ${\langle}P_{b'}(x)|P_b(x)\rangle=\delta_{b'b}$ and $\sum_b |P_b(x)\rangle{\langle} P_b(x)|=1$. Here we are omitting $|_x$ in the momentum eigenstates. Note that just as time-dependent potentials in the usual QM yield time-dependent energy eigenstates, a space-dependent potential in the STS extension leads to space-dependent momentum eigenstates. This means that the states with well-defined momentum depend on the position at which we describe the particle. In contrast, recall that momentum (energy) eigenfunctions in the usual QM (STS extension) are always the same, proportional to $\exp\{iPx/\hbar\}$ ($\exp\{-iEt/\hbar\}$).

Analogous to Eq.~(\ref{expansion}), from the linearity of the SC Shr\"odinger equation~(\ref{SchroT}), an arbitrary physical state in ${\cal H}|_x$ at a position $x$ can be expanded in the momentum basis as
\begin{equation}\label{expansionT}
    \ket{{\phi}(x)}=\sum_b~ {\tilde \phi}(b|x) \ket{P_b(x)},
\end{equation}
where
\begin{equation}\label{phiP}
   {\tilde \phi}(b|x)\equiv \langle P_b(x)|{\phi}(x)\rangle= C(b|x)e^{i\theta_b(x)},
\end{equation}
with $\theta_b(x)=1/\hbar\int_0^x dx' P_b(x')$ is the equivalent to the dynamical phase $\theta_a(t)$ in the usual QM. From the SC Schr\"odinger equation~(\ref{SchroT}), we can also readily verify that the coefficient $C_b(x)$ satisfies an equation similar to~(\ref{Ca}),
\begin{equation}
    \frac{dC(b|x)}{dx}=-\sum_{b'} C(b'|x)e^{i\Delta \theta_{b'b}(x)}\bra{P_b(x)}\frac{d}{dx}\ket{P_{b'}(x)},
\end{equation}
where $\Delta \theta_{b'b}(x)=\theta_{b'}(x)-\theta_b(x)$. Now, we have
\begin{equation}\label{prob}
{\cal P}(b|x)=\frac{|{\tilde \phi}(b|x)|^2}{{\langle {\phi}(x)|{ \phi}(x) \rangle}}
\end{equation}
as the probability amplitude of measuring the particle in the state $|P_b(x)\rangle$ (with momentum $P_{b}(x)$), given that the measurement happens at position $x$. Projecting Eq.~(\ref{expansionT}) on $|t\rangle$, we obtain an expansion equivalent to Eq.~(\ref{SolX3}),
\begin{equation}\label{SolT3}
{\pmb \phi}(t|x)=\langle t|{\phi}(x)\rangle=\sum_b ~{\tilde \phi}(b|x)~ {\pmb \phi}_b(t|x),
\end{equation}
where  
\begin{equation}\label{phib}
   {\pmb \phi}_b(t|x)=\mqty(\langle t|P^+_b(x)\rangle  \\ \langle t|P^-_b(x)\rangle)\equiv \mqty( \phi^+_b(t|x)   \\ \phi^-_b(t|x) )
\end{equation}
is the probability amplitude for the particle to arrive at time $t$, given that the detector is at position $x$ and its state is $|P_b(x)\rangle$. Now, using $\mathbbm {\hat P}$ defined in Eq.~(\ref{ruleP}), the projection of Eq.~(\ref{SchroTP1}) on $|t\rangle$ gives
\begin{equation}\label{SchroTP}
\sigma_z \sqrt{2m\left(i\hbar \pdv{t}-V(x,t)\right)}{\pmb \phi}_b(t|x)=P_b(x){\pmb \phi}_b(t|x).
\end{equation}

Reference~\cite{Dias} analyzed the momentum eigenvalue equation~(\ref{SchroTP}) for the free particle situation, where $[{\hat {\mathbbm P}},{\hat {\mathbbm H}}]=0$. Similar to the time-independence in the usual QM, the momentum eigenstate~(\ref{Pb}) becomes space-independent, $|P_b(x)\rangle=|P\rangle|_x$, where
\begin{equation}\label{SolTL1}
|P\rangle|_x=\ket{\phi_P}|_x \otimes |+\rangle=\mqty(\ket{\phi_P}|_x \\ 0)~~~{\rm for}~P>0
\end{equation}
and
\begin{equation}\label{SolTL2}
|P\rangle|_x=|\phi_{|P|}\rangle|_x \otimes |-\rangle=\mqty(0 \\ |\phi_{|P|}\rangle|_x)~~~{\rm for}~P<0,
\end{equation}
with $P_b(x)=\pm |P|=\pm \sqrt{2mE}$. Also, ${\tilde \phi}(b|x)$ defined in Eq.~(\ref{phiP}) can be written as
\begin{eqnarray}\label{phiPFree}
  {\tilde \phi}(P|x)={\tilde \phi}(P)e^{iP x/\hbar}.
\end{eqnarray} 
Thus, the general solution~(\ref{expansionT}) becomes
\begin{equation}\label{expansionTL}
\begin{aligned}
|\phi(x)\rangle &=\int_{-\infty}^\infty dP ~ {\tilde \phi}(P) ~e^{iPx/\hbar}~|P\rangle|_x \\
&=\int_0^\infty dP \mqty({\tilde \phi}(P) ~e^{iPx/\hbar}~\ket{\phi_P}|_x \\ {\tilde \phi}(-P) ~e^{-iPx/\hbar} ~ \ket{\phi_P}|_x),
\end{aligned}
\end{equation}
where we consider $E>0$. In the next section, we will discuss the physical difference between $|P\rangle|_x$ and $|P\rangle|_t$.

Under the circumstances above, the momentum eigenfunction ${\pmb \phi}_b(t|x)$ in Eq.~(\ref{phib}) becomes space-independent (${\pmb \phi}_b(t|x)={\pmb \phi}_P(t)=\langle t|P\rangle|_x$) and is given by
\begin{equation}\label{SolTL1}
    {\pmb \phi}_{P}(t)=\mqty(\langle t|\phi_P\rangle|_x \\ 0)=\mqty(\phi_{P}(t)   \\  0)~~~{\rm for}~P>0,
\end{equation}
and
\begin{equation}\label{SolTL2}
    {\pmb \phi}_{P}(t)=\mqty(0 \\ \langle t|\phi_{|P|}\rangle|_x )=\mqty(0 \\ \phi_{|P|}(t) )~~~{\rm for}~P<0,
\end{equation}
with
\begin{eqnarray}\label{SolTFree}
\phi_{|P|}(t)=\sqrt{\frac{|P|}{2\pi
m\hbar}}~e^{-iP^2t/(2m\hbar)}.
\end{eqnarray}
Finally, projecting Eq.~(\ref{expansionTL}) on $|t\rangle$, the SC wave function~(\ref{SolT3}) becomes
\begin{eqnarray}\label{SolTL}
    {\pmb \phi}(t|x)=\int_0^\infty dP \mqty({\tilde \phi}(P) ~e^{iPx/\hbar}~\phi_{P}(t) \\ {\tilde \phi}(-P)~ e^{-iPx/\hbar} ~ \phi_{P}(t)),
\end{eqnarray}
where its modulus square is ${\cal P}(t|x)$ given in Eq.~(\ref{pd}). It is worth noticing the similarity between Eqs.~(\ref{SchroXE})-(\ref{SolXFree}) and Eqs.~(\ref{SchroTP})-(\ref{SolTL}). Nevertheless, in contrast to the usual QM, where ${\cal P}(E;\pm|t)=|{\bar \psi}(E;\pm)|^2$ is time-independent for free particles, ${\cal P}(P|x)=|{\tilde \phi}(P)|^2/\langle {\phi}(x)|{\phi}(x)\rangle$ [see Eq.~(\ref{prob})] can depend on $x$ because of the normalization factor $\langle {\phi}(x)|{\phi}(x)\rangle$.

\section{An interpretation of the STS extension and its connection to QM}\label{interpretation}

First, it should pointed out that Ref.~\cite{Dias} primarily focuses on the free particle case, which limits significantly  the physical consequences predicted by the STS extension. Moreover, Ref. \cite{Dias} also concentrates on arrival time probability amplitudes (specifically, $\phi(t|x)$ and ${\phi_P}(t)\equiv \phi(t|P)$) and does not provide a proper interpretation of the wave functions of other observables of the particle. In particular, when the momentum probability amplitude ${\tilde \phi}(P)$ is introduced, it is identified, without physical arguments, as the momentum probability amplitude of the standard QM. As a result, Ref.~\cite{Dias} obtains the Kijowski distribution as the solution of the SC Schrödinger equation for the free particle case (as discussed at the end of Sec.~\ref{review}).

In contrast, in this section we will give an interpretation of the STS extension that challenges the connection between $|\phi(t|x)|^2$ and the Kijowski distribution established in Ref.~\cite{Dias}. We will begin by interpreting the solutions of the SC Schr\"odinger equation and attributing physical meaning to the eigenstates (along with their wave functions) of any observable in the STS extension. Thus, we will observe that the differences in the measurement procedures for obtaining ${\tilde \phi}(P)$ and ${\tilde \psi}(P)$ strongly indicate that these wave functions in general diverge from each other. After comparing the information contained in $|\psi(t)\rangle$ and $|{\phi}(x)\rangle$ using the equations of Sec.~\ref{comparison}, we will conclude that they provide complementary information about the same particle.

To propose an interpretation of the solutions of the SC Schrödinger equation~(\ref{SchroT}), let us keep in mind the information provided by the time-dependent Schrödinger equation: Knowing the initial wave function $\psi(x|t_0)$ --- the probability amplitude of finding the particle at position $x$, given that the observation takes place at the initial time $t_0$ --- the solution of the Schr\"odinger equation, $\psi(x|t)$, provides the probability amplitude of finding the particle at $x$, given that the observation takes place at a later time $t>t_0$. Note that time is an external classical parameter (the time of the laboratory clock) and hence can be chosen with arbitrary precision to evaluate the state of the system and its probabilities.

\begin{figure}
\includegraphics[width=0.5\textwidth]{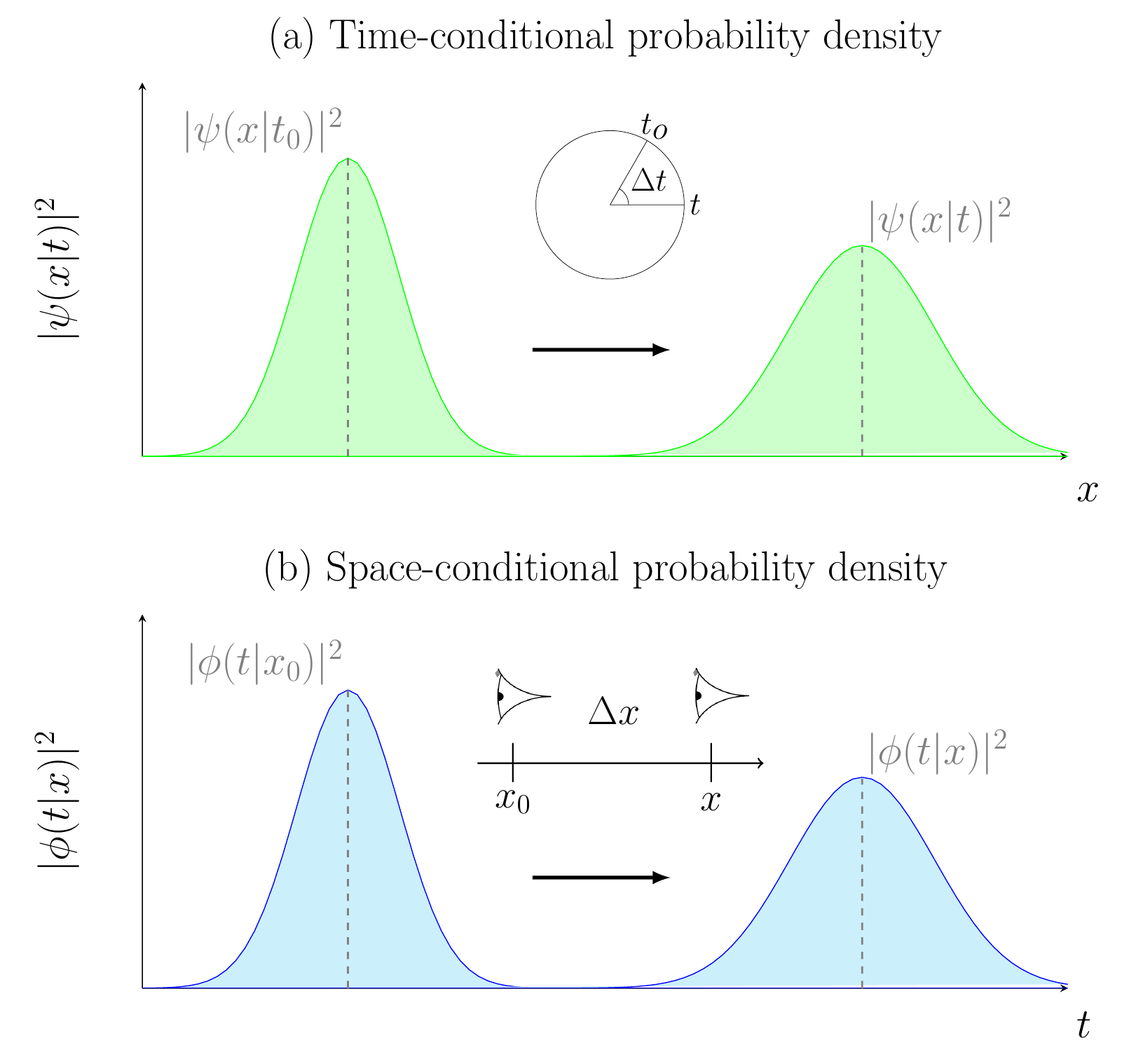}
\caption{Illustrations of the time (panel (a)) and space (panel (b)) translations prescribed by the Schr\"odinger and the SC Schr\"odinger equations, respectively. Knowing $\psi(x|t_0)$ ($\phi(t|x_0)$), the (SC) Schr\"odinger equation provides $\psi(x|t)$ ($\phi(t|x)$).} 
\label{Illustration}
\end{figure}

As discussed in the previous section, the STS extension is formulated by switching the roles of position and time (and energy and momentum) in QM. Therefore, following the same reasoning as in the previous paragraph and taking $x \rightleftarrows t$, the interpretation of the solutions of the SC Schr\"odinger equation~(\ref{Schro2T}) becomes the following: Knowing the ``initial'' SC wave function ${\pmb \phi}(t|x_0)$ --- the probability amplitude for the particle to arrive at instant $t$, given that the observation happens at position $x_0$ --- the solution of Eq.~(\ref{Schro2T}), ${\pmb \phi}(t|x)$, provides the probability amplitude of the particle arriving at $t$, given that the observation occurs at $x \lessgtr x_0$. Now, position is the external parameter (the position of the classical detector). To better understand this interpretation, Fig.~\ref{Illustration} illustrates the difference between the time evolution of QM and the ``space'' evolution of its STS extension.

Analogous to the standard QM, this interpretation works for any observable of the particle, not only the time of arrival. In practice, by knowing $|\psi(t_0)\rangle$, we are able to predict measurements of observables of the particle performed by a detector designed to measure precisely at time $t_0$ and large enough to interact with the whole wave packet, $\psi(x|t_0)$, at $t_0$. The necessity of a spatially extended detector will be discussed later in the context of momentum measurements. Then, by providing $|\psi(t)\rangle$, the Schr\"odinger equation enables us to predict measurements when the same detector is designed to measure at time $t>t_0$. In a similar manner, by knowing $|\phi(x_0)\rangle$, we are able to predict measurements of the same observables as the usual QM, but now performed by a detector localized at position $x_0$ and ``activated'' all the time. Then, by providing $|\phi(x)\rangle$, the SC Schr\"odinger equation enables us to predict measurements when the same detector is localized at position $x \lessgtr x_0$.

To make this interpretation more transparent, let us express the ``space evolution'' of $|\phi(x)\rangle$ in a manner analogous to the time evolution of $|\psi(t)\rangle$, in which we use the time-evolution operator. Given a state $\ket{\phi(x_0)}$, we define
\begin{equation}\label{spaceoperator}
    \ket{\phi(x)}=\hat{\mathbbm U}(x,x_0)\ket{\phi(x_0)},
\end{equation}
where $\hat{\mathbbm U}(x,x_0)$ is the space-evolution operator. Plugging Eq.~(\ref{spaceoperator}) into the SC Schr\"odinger equation~(\ref{SchroT}), we obtain
\begin{equation}\label{SOequation}
   \hat{\mathbbm{P}}(x) \hat{\mathbbm U}(x,x_0)= -i\hbar \frac{d}{dx} \hat{\mathbbm U}(x,x_0).
\end{equation}

Note that, just as in the standard QM, where it is possible to have $[{\hat {\rm H}}(t_1),{\hat {\rm H}}(t_2)]\neq 0$, here, we can have $[\hat{\mathbbm P}(x_1),\hat{\mathbbm P}(x_2)]\neq 0$. Therefore, to solve Eq.~(\ref{SOequation}), we first introduce a space-ordering operator, analogous to the time-ordering operator in the usual QM,
\begin{equation}\label{spaceordering}
    \hat{\cal{S}}_{\rightleftarrows}
(\hat{\mathbbm A}(x_1)\hat{\mathbbm B}(x_2))=
     \begin{cases}
        \hat{\mathbbm B}(x_2)\hat{\mathbbm A}(x_1), & \text{if } x_1 \lessgtr x_2\\
       \hat{\mathbbm A}(x_1)\hat{\mathbbm B}(x_2), & \text{if } x_2\lessgtr x_1,
    \end{cases}
\end{equation}
where $\hat{\cal{S}}{\rightarrow}$ ($\hat{\cal{S}}_{\leftarrow}$) orders operations in terms of increasing (decreasing) values of $x$. The need for the two ordering operators, $\hat{\cal{S}}_{\rightleftarrows}$, arises from the fact that, unlike in the standard QM, where the temporal evolution has a single direction (forward in time), in the STS extension, we can ``evolve'' back and forth in space. In this picture, the general solution of Eq.~(\ref{SOequation}) is given by
\begin{equation}\label{solutionU}
    \hat{\mathbbm U}_{\rightleftarrows}(x,x_0)=\hat{\cal S}_{\rightleftarrows} \bigg[e^{\frac{i}{\hbar} \bigintssss_{x_0}^{x} d{x'} \hat{\mathbbm{P}}(x')} \bigg],
\end{equation}
where we must use  $\hat{\mathbbm U}_{\rightarrow}(x,x_0)$ for $x>x_0$ and $\hat{\mathbbm U}_{\leftarrow}(x,x_0)$ for $x<x_0$. It is worth noting that since $\hat{\mathbbm{P}}(x)$ is not always Hermitian, which occurs when $V(x)>E$, the space-evolution operator is not necessarily unitary, as expected. Thus, the norm of $\ket{\phi(x_0)}$ may not be preserved when translating it to $\ket{\phi(x)}$. For $\hat{\mathbbm U}_{\rightarrow}(x,x_0)$ ($\hat{\mathbbm U}_{\leftarrow}(x,x_0)$) to be unitary, we must have $E>V(x)$ throughout the entire interval $[x_0,x]$ ($[x,x_0]$). 


At first sight, one can think the STS extension is not designed to answer the ordinary TOA problem: Given a particle with a wave function of positive momenta at $t_0$, $\psi(x|t_0)$, restricted to a region $x<x^*$, when does this particle arrive at $x^*$? This is because, following the above interpretation, just as the Schr\"odinger equation provides time translations of a previously known wavefunction $\psi(x|t_0)$, the SC Schr\"odinger equation~(\ref{Schro2T}) describes space translations of a SC wavefunction ${\pmb \phi}(t|x_0)$ that should also be known. Besides, we cannot figure out a certain conditional distribution ${\cal P}(a|b)$ ($|\phi(t|x_0)|^2$) only by knowing ${\cal P}(b|a)$ ($|\psi(x|t_0)|^2$). Nevertheless, if $\psi(x|t)$ and ${\pmb \phi}(t|x)$ share common information about other observables, e.g., energy and/or momentum, $\psi(x|t_0)$ can eventually determine ${\pmb \phi}(t|x_0)$ and vice-versa.

With this discussion in mind, let us compare $|\psi(t)\rangle$ and $|{\phi}(x)\rangle$ more deeply by representing them in terms of eigenstates associated with the same observable. We choose the momentum eigenstates and compare Eq.~(\ref{expansionT}) with
\begin{equation}\label{SolXP}
|\psi(t)\rangle=\int_{-\infty}^{\infty} dP ~ {\tilde \psi}(P|t) |P\rangle|_t,
\end{equation}
where $|P\rangle|_t=1/\sqrt{2\pi \hbar}~\int_{-\infty}^{\infty}dx~e^{iPx/\hbar}|x\rangle$. Let us begin the comparison by proposing the physical differences between $|P\rangle|_t$ and $|P_b(x)\rangle$.

When one says, e.g., that a system has a well-defined momentum in usual QM, implicitly, one is claiming that $|\psi(t)\rangle=|P\rangle|_t$, i.e., the particle has momentum $P$ (and indefinite position) at instant $t$. Therefore, eigenstates in QM refer to a well-defined property at a fixed time, justifying the notation $|\rangle|_t$. On the other hand, based on the interpretation of the SC Schr\"odinger equation, we interpret $|\phi(x)\rangle=|P_b(x)\rangle$ ($|P\rangle|_x$ for the free particle situation) as the state of a particle with momentum $P_b(x)$ at position $x$, i.e., the particle arrives at $x$ (at an indefinite time) with momentum $P_b(x)$.

The difference between $|P\rangle|_t$ and $|P_b(x)\rangle|_x$ also becomes evident by noting that $|P\rangle|_t$ is the Fourier transform of $|x\rangle$, which involves all space, and has the same expression regardless of the applied potential. In contrast, $|P_b(x)\rangle$ is defined at a specific position and is the solution of the momentum eigenvalue equation~(\ref{SchroTP1}), thus depending on the potential involved.

Keeping in mind the physical distinction between $|P\rangle|_t$ and $|P_b(x)\rangle$, we now focus on comparing the momentum wave functions ${\tilde \psi}(P|t)$ and ${\tilde \phi}(b|x)$. When incorporating the interpretations above into the state of the particle $|\phi(x)\rangle$, the wave function ${\tilde \phi}(b|x)$ becomes the probability amplitude of the particle arriving at position $x$ with momentum $P_b(x)$. Note that similar to the space-independence of $|{\tilde \phi}(b|x)|^2=|{\tilde \phi}(P|x)|^2$ in Eq.~(\ref{phiPFree}),
$|{\tilde \psi}(P|t)|^2$ --- the probability density of measuring the particle with $P$, given that the observation happens at $t$ --- is time-independent for the free particle situation,
\begin{equation}\label{SolXP2}
{\tilde \psi}(P|t)={\tilde \psi}(P)~e^{-i P^2t/(2m\hbar)}.
\end{equation}

In general, the impossibility of connecting these two wave functions is because while $|{\tilde \psi} (P|t)|^2$ predicts experimental data of momentum collected at a fixed instant $t$, $|{\tilde \phi}(b|x)|^2$ predicts experimental data of momentum collected at a fixed position $x$. In this scenario, as discussed above, to obtain the predictions of a time-dependent $|{\tilde \psi} (P|t)|^2$ accurately, the detector must be big enough to interact simultaneously with the entire wave packet $\psi(x|t)$ precisely at time $t$. This is because  ${\tilde \psi}(P|t)$ is the Fourier transform of $\psi(x|t)$, which is a feature of the whole wave function at $t$. In contrast, the predictions of ${\tilde \phi}(b|x)$ needs for the presence of a detector located at position $x$ that does not select any time $t$.

An example of this incompatibility is the following. If a wave function $\psi(x|t)$ cannot cross a potential barrier, the particle never reaches a certain position $x^*$ on the transmission side. In this region, the particle is free and ${\tilde \phi}(b|x^*)={\tilde \phi}(P|x^*)$ is zero for all $P$, while ${\tilde \psi}(P|t) \neq 0$. We see that linking the momentum wave functions leads to a similar problem to relating $\psi(x|t)$ and ${\pmb \phi}(t|x)$; while $|\psi(t)\rangle$ describes observables of the particle at a given time, $|{\phi}(x)\rangle$ describes the same observables but at a given position.

Let us focus the discussion above on the simplest possible situation, a free particle with positive momenta that always arrives at a given point $x^*$, i.e., $\langle {\phi}(x^*)|{\phi}(x^*)\rangle=1$. This situation is described by the solution~(\ref{SolTL}) with ${\tilde \phi}(-P)=0$. As the entire wave packet $\psi(x|t)$ passes through $x^*$, all possible momenta of the time-independent distribution $|{\tilde \psi}(P|t)|^2=|{\tilde \psi}(P)|^2$ can be measured if a detector is at $x^*$. In this scenario, one may expect that $|{\tilde \psi}(P)|^2$ equals $|{\tilde \phi}(P|x^*)|^2=|{\tilde \phi}(P)|^2$ (the probability for the particle to arrive $x$ with momentum $P$, regardless of the time it arrives). If one also assumes their phases are the same, i.e., ${\tilde \phi}(P)={\tilde \psi}(P)$ (which is not a trivial statement), ${\cal P}(t|x)$ of Eq.~(\ref{pd}) becomes the Kijowski distribution, as considered in Ref.~\cite{Dias} without further justification. 

An opposing viewpoint to the validity of $|{\tilde \psi}(P)|^2=|{\tilde \phi}(P)|^2$ is as follows. We are aware that to measure $|\phi(P|x)|^2=|{\tilde \phi}(P)|^2$, the detector should be situated at position $x$. Describing this measurement within the standard QM, even for a time-independent $|{\tilde \psi}(P|t)|^2=|{\tilde \psi}(P)|^2$, we envision the Schrödinger wave function of the particle gradually reaching (and interacting with) the detector, starting with its front tail. Consequently, one may expect that the momentum distribution measured by this localized detector diverges from $|{\tilde \psi}(P)|^2$, which should involve an interaction of the entire $\psi(x|t)$ simultaneously at time $t$, as previously discussed. Furthermore, it is important to note that assuming ${\tilde \phi}(P)={\tilde \psi}(P)$ is equivalent to assuming that $|P\rangle|_t$ and $|P\rangle|_x$ represent the same quantum state, which is not true according to our interpretation of the STS extension.

In this context, in scenarios where the Schrödinger packet is spatially narrow and fast enough, we may have $|{\tilde \psi}(P)|^2 \approx|{\tilde \phi}(P)|^2$. Nevertheless, the eventual validity of the relationship $|{\tilde \psi}(P)|^2 \approx|{\tilde \phi}(P)|^2$ in some physical regime can only be established with a detailed description of the experimental procedure for obtaining $|{\tilde \phi}(P|x)|^2$. Note that merely stating the need for a localized detector is insufficient. Just as challenges arise when measuring the particle's arrival time (including issues like particle reflection), similar challenges may occur when measuring the particle momentum with the detector at a fixed position.

With this discussion in mind, it should be pointed out that as is the goal of any ideal TOA model, we expect that the predictions of $|{\pmb \phi}(t|x)|^2$ can be confirmed by taking some limits of ideal measurements, where well-designed detectors and/or clocks coupled to the particle register the TOA of a particle. Describing this situation via the usual QM, the information of $|{\pmb \phi}(t|x)|^2$ should be contained, e.g., in the clock's state belonging to ${\cal H}|_t$, and not in the free-measurement state of the particle ($\psi(x|t)$). With this result, we can test, e.g., the validity of $|{\tilde \psi}(P)|^2 \approx|{\tilde \phi}(P)|^2$ in some physical regime. If this relation is valid and hence $|{\pmb \phi}(t|x)|^2$ is approximately the Kijowski distribution for the free particle case, it can be measured from the time probability density of detecting the first emitted photon when a two-level atom enters a laser-illuminated region~\cite{Damborenea}. By using the quantum jump technique and ``operator normalization''~\cite{Brunetti}, this probability density becomes the Kijowski distribution in the limit of a strong laser field and fast decay~\cite{Heger}.





From the discussion above, we verify that in general the information provided by the STS extension is not entirely embodied in the {\it intrinsic} state of the particle $|\psi(t)\rangle$ (a free-measurement state). Thus, $|\psi(t)\rangle$ and $|\phi(x)\rangle$ provide complementary information about the same particle, and as a result the quantum state of the particle at time $t$ is as fundamental as its quantum state at position $x$.


\section{Solutions for $V=V(x)$ and the TOA of a particle crossing a potential barrier}\label{solution}

In Sec.~\ref{GS}, we will solve the SC Schr\"odinger equation for an arbitrary time-independent potential. Then, in Sec.~\ref{comparison2}, we will apply this solution to predict the TOA of a particle traversing a square potential barrier. We will assume that ${\tilde \phi}(P)={\tilde \psi}(P)$ for a free incident particle with momentum $P>0$ and use this relation for the initial condition $\phi(t|x_0)$ on the left side of the barrier. We will compare our results with a generalization of the Kijowski distribution~\cite{Delgado,Leon,Baute2}. The consequences of considering ${\tilde \phi}(P)={\tilde \psi}(P)$ will be discussed. We will not consider the case of ${\tilde \phi}(P) \neq {\tilde \psi}(P)$ on the left side of the barrier because in this work, we will not propose an operational method to obtain or measure ${\tilde \phi}(P)$, making it challenging to justify any specific choice of ${\tilde \phi}(P)$ for the ``initial'' condition $\phi(t|x_0)$.

\subsection{General solution of the SC Schr\"odinger equation for $V=V(x)$}\label{GS}

It is convenient to solve the SC Schr\"odinger equation~(\ref{SchroT}) for an arbitrary time-independent potential $V=V(x)$ using the energy representation, $\{|E\rangle|_x\}$,
\begin{eqnarray}\label{SolTE}
|{\phi}(x)\rangle&=&\int_{-\infty}^{\infty} dE ~ {\bar {\pmb \phi}}(E|x) |E\rangle|_x\nonumber\\
&=&\int_{-\infty}^{\infty} dE \mqty({\bar \phi}^+(E|x)\\{\bar \phi}^-(E|x)) \ket{E}|_x,
\end{eqnarray}
where $\langle t|E\rangle|_x=1/\sqrt{2\pi \hbar}~e^{-iEt/\hbar}$ and $|{\bar {\pmb \phi}}(E|x)|^2={\bar {\pmb \phi}}^\dagger(E|x) {\bar {\pmb \phi}}(E|x)$ is the probability density for the particle to arrive at $x$ with energy $E$, regardless of its arrival time. For the free particle situation~\cite{Dias},
\begin{equation}\label{SolTE2}
{\bar \phi}^{\pm}(E|x)={\bar \phi}^{\pm}(E)~e^{\pm i \sqrt{2mE} x/\hbar}.
\end{equation}

Substituting Eq.~(\ref{SolTE}) into the SC Schr\"odinger equation~(\ref{SchroT}), we have
\begin{eqnarray}\label{sol1}
&&\int_{-\infty}^{\infty} dE~{\sigma}_{z}~{\bar {\pmb \phi}}(E|x) ~\sqrt{2m\big[{\hat {\mathbbm H}}-V(x){\hat {\mathbbm 1}}\big]}~|E\rangle|_x\nonumber\\
&=&-\int_{-\infty}^{\infty} dE~i\hbar\frac{\partial {\bar {\pmb \phi}}(E|x)}{\partial x}~|E\rangle|_x.
\end{eqnarray}
Expanding the operator $\sqrt{{\hat {\mathbbm H}- V(x){\hat {\mathbbm 1}}}}$, for $V(x)\neq 0$, in power series of $\mathbbm H$, we obtain
\begin{eqnarray}
\sqrt{{\hat {\mathbbm H}}-V(x){\hat {\mathbbm 1}}}=\sum_{n=0}^{\infty} {\frac{1}{2}\choose n} i^{1-2n} [V(x)]^{1/2-n}~{\hat {\mathbbm H}}^n,
\end{eqnarray}
which applied to $|E\rangle|_x$ yields
\begin{eqnarray}
&&\sum_{n=0}^{\infty} {\frac{1}{2}\choose n} i^{1-2n} [V(x)]^{1/2-n}~{\hat {\mathbbm H}}^n|E\rangle|_x\nonumber\\
&=&\sum_{n=0}^{\infty} {\frac{1}{2}\choose n} i^{1-2n} [V(x)]^{1/2-n}E^n|E\rangle|_x\nonumber\\
&=&\sqrt{E-V(x)}~|E\rangle|_x.
\end{eqnarray}
Substituting this equation back into Eq.~(\ref{sol1}), and projecting the resulting expression on $|E'\rangle|_x$, we obtain a differential equation for ${\bar {\pmb \phi}}(E'|x)$ given by
\begin{eqnarray}
{\sigma}_z\sqrt{2m\big[E'-V(x)\big]}~{\bar {\pmb \phi}}(E'|x)=-i\hbar\frac{\partial}{\partial x}{\bar {\pmb \phi}}(E'|x).
\end{eqnarray}
The solution for each component is
\begin{equation}\label{SolE}
{\bar \phi}^\pm(E|x)={\bar \phi}^\pm(E|x_0) e^{{\pm}i \bigintssss_{x_0}^x dx'\sqrt{2m [E-V(x') ]}/\hbar},
\end{equation}
where we replaced $E'$ with $E$. Substituting Eq.~(\ref{SolE}) into $|{ \phi}(x)\rangle$ of Eq.~(\ref{SolTE}), and projecting the resulting expression on $|t\rangle$ yields
\begin{eqnarray}\label{SolTX}
&&\phi^\pm(t|x)=\nonumber\\
&&\int_{-\infty}^{\infty}dE~\frac{{\bar \phi}^\pm(E|x_0)}{\sqrt{2\pi \hbar}}~e^{{\pm}i \bigintssss_{x_0}^x dx'\sqrt{2m [E-V(x') ]}/\hbar-iEt/\hbar}.\nonumber\\
\end{eqnarray}
This is the general solution for the probability amplitude of the ideal TOA at position $x$ of a particle under the action of $V=V(x)$. By taking $V(x)=0$, imposing normalization in time, and changing the variable of integration to $P$, we recover the free particle solution of Eq.~(\ref{SolTL}).

By inspecting Eq.~(\ref{SolTX}), it can be noticed that if, e.g., $V(x)>E$ in the interval $-\infty<x<x_0=0$, the solution $\phi^+(t| x)$ of Eq.~(\ref{SolTX}) diverges when $x \rightarrow -\infty$ as the integral in $x$ goes to $-i\infty$. To understand this non-physical behavior, first recall that since $\phi^+(t|x_0)$ describes the particle arriving at $x=x_0=0$ from the left, $\lim_{x\rightarrow -\infty}\phi^+(t |x)$ describes the particle arriving at $x\rightarrow -\infty$ from the left. Thus, if $\phi^+(t|x_0)$ is finite, the divergence of $\lim_{x\rightarrow -\infty}\phi^+(t|x)$ means that it is impossible for a particle to start from $x\rightarrow -\infty$ and arrive at $x_0 $ with a finite probability when, along its entire path, its energy is lower than the applied potential.

It is worth noting that, for the same potential considered above ($V(x)>E$ in $[-\infty,0]$), the aforementioned divergence can be eliminated by taking $x_0 \rightarrow -\infty$. In this case, when normalizing $\phi^+(t|x_0)$, the general solution~(\ref{SolTX}) yields $\phi^+(t|x=0) \approx 0$. This result agrees with the previous analysis, indicating once again that a particle starting from $x \rightarrow -\infty$ does not arrive at $x_0$ (as $|\phi^+(t|x=0)|^2 \approx 0$) if $V(x)>E$ over a sufficiently long region along its trajectory. This behavior is expected as in the standard QM, the Schr\"odinger wave function decays exponentially when the particle enters a region with $V(x)>E$, eliminating any probability of finding it in deeper regions of the potential.

We conclude this subsection pointing out the similarity between Eq.~(\ref{SolTX}), which is valid for $V=V(x)$, and the solution of the Schr\"odinger equation for potentials that depend exclusively on time, $V=V(t)$, which is given by
\begin{equation}\label{SolXT}
\psi(x|t)=\int_{-\infty}^{\infty} dP \frac{{\tilde \psi}(P|t_0)}{\sqrt{2\pi \hbar}} e^{-i\bigintssss_{t_0}^t dt'[{P^2}/(2m)+V(t')]/\hbar +iPx/\hbar}.
\end{equation}
Here we observe that one can go from one solution to the other via the transformation $(t,P,H(t)) \rightarrow (x,E,\pm P(x))$, with $H(t)$ and $P(x)$ given by the classical expressions of Eqs.~(\ref{ruleX}) and~(\ref{ruleP}), respectively. This symmetry becomes even more apparent by identifying the classical action $S$ in Eq.~(\ref{SolTX}), for $V=V(x)$, and in Eq.~(\ref{SolXT}), for $V=V(t)$, which allows us to rewrite them as
\begin{equation}
\phi^\pm(t|x)=\frac{1}{\sqrt{2\pi \hbar}}\int_{-\infty}^{\infty} dE~ {\bar \phi}^\pm(E|x_0)~e^{-iS(E,x)/\hbar}
\end{equation}
and
\begin{equation}
\psi(x|t)=\frac{1}{\sqrt{2\pi \hbar}}\int_{-\infty}^{\infty}d{P}~ {\tilde \psi}(P|t_0)~e^{-iS(P,t)/\hbar}.
\end{equation}

Using the solution of Eq.~(\ref{SolTX}), in the next subsection, we will calculate the TOA distribution of a particle crossing a potential barrier.

\subsection{Comparing the STS extension with a generalization of the Kijowski distribution}
\label{comparison2}

Let us apply the solution of Eq.~(\ref{SolTX}) to a free particle prepared far to the left and detected after crossing a square barrier of height $V_0$, width $L$, and located in the interval $0 < x < L$. The particle initially ($t_i=0$) has a state in the usual QM given by a Gaussian wave packet centered at $x_i$ ($\hbar=1$), 
\begin{equation}\label{initialX}
\psi(x|0)=\frac{1}{(2\pi \delta^2)^{1/4}}e^{- \left[(x-x_i)/(2\delta)-iP_i\delta\right]^2- P_i^2\delta^2 },
\end{equation}
where $\psi(x|0)=\langle x|\psi(0)\rangle$. Here $x_i$, $P_i$, and $\delta$ will assume values such that the packet is to the left of the origin and has only positive momenta.

To compute the TOA on the transmission side using the interpretation of the SC Schr\"odinger equation given in the previous section, we should apply Eq.~(\ref{SolTX}) to an initial probability amplitude of the TOA, ${\pmb \phi}(t|x_0)$. If we want the solution, ${\pmb \phi}(t|x>L)$, to take into account the barrier traversal time, $x_0$ must be located on the left side of the barrier, $x_i < x_0 \leq 0$. We consider $\psi(x_0|t_i)\approx 0$, meaning that at time $t_i$, the particle has not arrived at $x_0$. As discussed in Sec.~\ref{interpretation}, since the particle travels freely on the left side of the barrier and it always passes through the point $x_0$, let us assume that $|{\pmb \phi}(t|x_0)|^2$ is the Kijowski distribution~(\ref{pd}), i.e., ${\tilde \phi}(P)={\tilde \psi}(P)$, where ${\tilde \psi}(P)$ is the momentum wave function of $\psi(x|t_i)$,
\begin{equation}\label{initialXP}
{\tilde \psi(P)}=\left(\frac{2\delta^2}{\pi}\right)^{1/4}e^{-\delta^2 (P-P_i)^2 -iPx_i}.
\end{equation}

As we are interested in the TOA of the particle in the transmitted region, where there are only positive momenta, we will focus exclusively on $\phi^+(t|x)$. Using Eqs.~(\ref{SolTL}) and~(\ref{initialXP}) under the circumstances above, the ``initial'' condition at $x_0$ of the SC wave function becomes 
\begin{eqnarray} \label{initialT}
\phi^+(t|x_0)= 
\int_0^{\infty}dP~{\tilde \psi}(P)\sqrt{\frac{P}{2\pi
m\hbar}}~e^{iPx_0/\hbar-iP^2t/(2m\hbar)},\nonumber\\
\end{eqnarray}
where $P=\sqrt{2mE}$. To apply the SC solution~(\ref{SolTX}) to $\phi^+(t|x_0)$, Eq.~(\ref{initialT}), we have to figure out ${{\bar \phi}^+}(E|x_0)$ for this particular ``initial'' condition. Considering $x_0=0$, Eq.~(\ref{SolTX}) at $x=x_0$, where $V(x_0)=0$, reduces to
\begin{eqnarray}\label{SolTX2}
\phi^+(t|x_0)=\frac{1}{\sqrt{2\pi \hbar}}
\int_{-\infty}^{\infty}dE~{\bar \phi}^+(E|x_0)e^{-iEt/{\hbar}}.
\end{eqnarray}
Changing the variable of integration $P$ in Eq.~(\ref{initialT}) to $E$, and comparing the resulting expression with Eq.~(\ref{SolTX2}), we identify
\begin{eqnarray}\label{Const}
{\bar \phi}^+(E|x_0)=\Theta(E)\left(\frac{m}{2E}\right)^{1/4} {\tilde \psi}(\sqrt{2mE}),
\end{eqnarray}
where $\Theta(E)$ is the Heaviside step function. Now, we have to substitute Eq.~(\ref{Const}) into the SC wave function~(\ref{SolTX}), considering $V(x)$ as the square potential barrier defined at the beginning of this section. Finally, evaluating the resulting expression at $x>L$, we obtain
\begin{eqnarray}\label{SolTBarrier}
\phi^+(t|x)&=&\frac{1}{\sqrt{2\pi \hbar}}
\int_{-\infty}^{\infty}dE  ~\Theta(E)\left(\frac{m}{2E}\right)^{1/4} {\tilde \psi}(\sqrt{2mE})\nonumber\\ &\times& e^{i\sqrt{ 2m \left( E - V_0 \right)} L/\hbar + i\sqrt{ 2m E} (x-L)/\hbar -iEt/\hbar}.\nonumber\\
\end{eqnarray}
This solution gives the time probability amplitude for the particle to arrive at $x>L$.

Considering ${\tilde \phi}(P)={\tilde \psi}(P)$ for the incident-free particle allows us to compute the arrival time after the barrier using the STS extension in another way. As the transmitted particle is free and always passes through $x>L$, one can also consider the free particle solution of Eq.~(\ref{SolTL}) (whose squared modulus is given by Eq.~(\ref{pd})) to describe the particle in this region, but now with amplitude ${\tilde \phi}(P)={\tilde \psi}_{T}(P)$, where ${\tilde \psi}_T(P)=T(P){\tilde \psi}(P)$ is the momentum wave function of the transmitted packet. Here,
\begin{equation}\label{Trans}
    T(P) = \frac{4 P P' e^{-i(P - P')L/\hbar}}{\left(P + P' \right)^2 - e^{2 iP'L/\hbar} \left(P - P' \right)^2}
\end{equation}
is the transmission amplitude and $P'=\sqrt{P^2-2mV_0}$. Replacing ${\tilde \phi}(P)$ with ${\tilde \psi}_{T}(P)$ in Eq.~(\ref{pd}), and taking ${\tilde \phi}(-P)=0$, ${\cal P}(t|x)$ becomes
\begin{equation}\label{KijoG}
    \Pi_{K}^{N} (t|x) =\frac{\left| \bigintsss_0^\infty dP~T(P){\tilde \psi}(P) \sqrt{P} e^{-i  P^2 t / (2m\hbar) + i  P x/\hbar}  \right|^2}{2 \pi m \hbar\bigintsss_0^\infty dP |T(P){\tilde \psi}(P)|^2}.
\end{equation}
Note that the probability for the particle to arrive at $x$, regardless of the time (${\langle { \phi}(x)|{ \phi}(x) \rangle}$) equals the probability for the particle to be transmitted ($\int_0^\infty dP |T(P){\tilde \psi}(P)|^2$). Equation~(\ref{KijoG}) is the normalized Kijowski distribution for the transmitted packet. This equation gives the probability density for the TOA at $x$, given that the particle has been transmitted through the potential barrier. Ximenes {\it et al.} considered this approach in Ref.~\cite{Ricardo}. We showed that for an electromagnetic experiment that simulates quantum tunneling~\cite{Ranfa}, the average traversal time obtained via Eq.~(\ref{KijoG}) agrees better with the experimental data than the traditional B\"uttiker-Landauer and phase-time models~\cite{ButLand,Buttiker,Wigner}.

It is worth pointing out that Eq.~(\ref{KijoG}) was obtained by different methods using the usual QM~\cite{Baute,Leon,Baute2,Heger2}. In particular, it arises from the same operational model discussed at the end of Sec.~\ref{interpretation}, which computes the detection time of the first emitted photon when a two-level atom enters a laser-illuminated region. In this situation, when one adds a square potential barrier and takes the limits of a strong laser field and fast decay, the TOA distribution at $x>L$ becomes Eq.~(\ref{KijoG}) for the transmitted particles~\cite{Heger2}. Below we compare the TOA distributions $\rho(t|x)\equiv {\cal P}(t|x)$, computed from Eq.~(\ref{SolTBarrier}), and $\Pi_K^N(t|x)$ of Eq.~(\ref{KijoG}). Thus, we investigate whether the assumptions ${\tilde \phi}(P)={\tilde \psi}(P)$ and ${\tilde \phi}_T(P)={\tilde \psi}_{T}(P)$ lead to inconsistencies in the STS extension.

\begin{figure}
\includegraphics[width=0.47\textwidth]{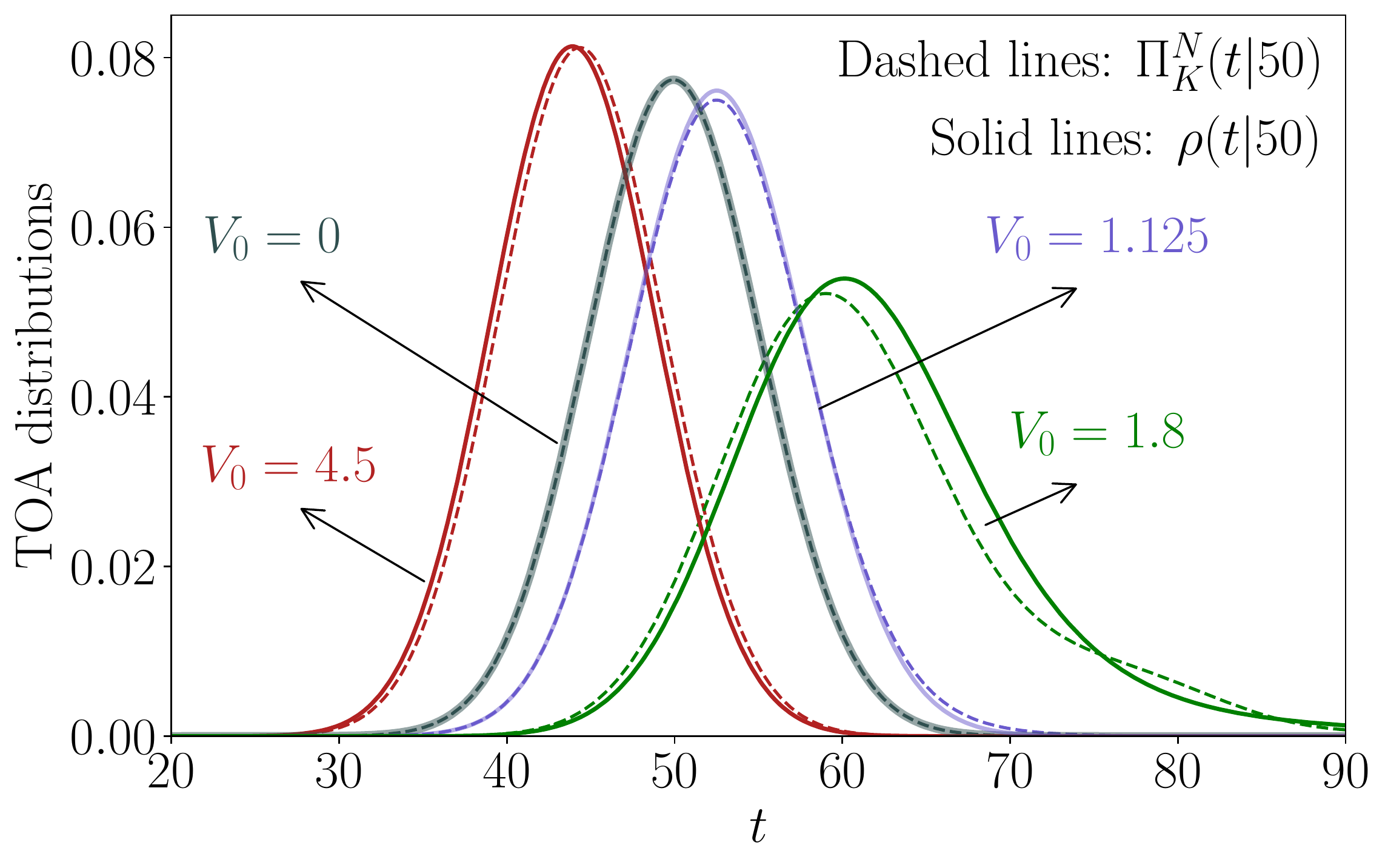}
\caption{Probability distributions for the arrival time of the transmitted particles at $x=50$. The initial wave packet $\psi(x,0)$ has $P_0=2$, $\delta=10$, $x_0=-50$, and $m=1$. The width of the barrier is $L = 10$. The solid (dashed) line illustrates the prediction of ${\cal P}(t|x)$ ($\Pi_K^N(t|x)$).}
\label{TOA}
\end{figure}

We plot the TOA probability distribution in Fig.~\ref{TOA} using physical parameters similar to those of Ref.~\cite{Leon}. This reference also uses the initial condition~(\ref{initialX}) and obtains Eq.~(\ref{KijoG}) by canonically transforming the quantum TOA of the free particle. The width of the barrier is $L=10$, the detector is at $x=50$, and the parameters of the initial wave packet~(\ref{initialX}) are $x_i=-50$, $P_0=2$, $\delta=10$, and $m=1$. Note that the average energy of the incident particle is $E_0=P_0^2/(2m)=2$. The solid and dashed curves in Fig.~\ref{TOA} show the TOA distribution at $x=50$ using Eqs.~(\ref{SolTBarrier}) and~(\ref{KijoG}), respectively, for $V_0=0$, $V_0=1.125$ ($<E_0$), $V_0=1.8$ ($<E_0$), and a tunneling regime with $V_0=4.5$ ($>E_0$). Note that the TOA of a free classical particle with velocity $P_0/m=2$ is $50$ since the total length of the particle's path is $100$ units.

Inspecting Fig.~\ref{TOA}, first, we observe that compared to the free particle situation, $V_0=0$, both solid and dashed curves illustrate a delay of the TOA for $V_0=1.125$ and $V_0=1.8$, and advancement for $V_0=4.5$. This advancement is responsible for the Hartman effect~\cite{Hartman}, as discussed in Ref.~\cite{Leon}. In addition, while the solution of the SC Schr\"odinger equation for the potential barrier, given by Eq.~(\ref{SolTBarrier}), and the Kijowski distribution~(\ref{KijoG}) are the same for $V_0=0$, they can predict different arrival times when $V_0 \neq 0$. In the tunneling regime, we verified that the larger the value of $V_0$, the closer the solid and dashed curves become, such that they are visually indistinguishable for $V_0 \gtrapprox 20$ in the scale of Fig.~\ref{TOA}. For more details about the predictions of Eq.~(\ref{KijoG}), see Ref.~\cite{Leon}.

The disagreement of Fig.~\ref{TOA} also shows that if both relations ${\tilde \phi}(P)={\tilde \psi}(P)$ and ${\tilde \phi}_T(P)={\tilde \psi}_{T}(P)$ are correct, which is not expected from the discussion of Sec.~\ref{interpretation}, some computation and/or interpretation of the STS extension must be somehow reformulated. Ultimately, the correct formalism (if any) of the STS extension will be defined by figuring out how $\phi(t|x)$ can actually be measured in the laboratory. A more elaborate investigation of the predictions of Eq.~(\ref{SolTX}) is currently being developed, including comparisons with other TOA approaches in the presence of interaction and the experiment of Ref.~\cite{Steinberg}. For example, Ref.~\cite{Heger2} obtains another generalization of the Kijowski distribution similar to Eq.~(\ref{KijoG}) but with ${\tilde \psi}_T(P)=T(P){\tilde \psi}(P)/|T(P)|$. We hope this analysis sheds light on the validity of the relationships used above.

\section{\label{sec:level7}Conclusion}


First, it is worth remarking that the STS extension has no intention to replace QM. It is in agreement with QM but enlarges the range of statistical scenarios to which QM is applicable. The most emblematic example refers to the ideal TOA, which, as discussed above, is still an open problem within the scope of traditional quantum theory.

In this work, we first derived in detail the STS extension of QM proposed in Ref.~\cite{Dias}. Then, considering arbitrary potentials, we investigated the momentum eigenvalue equation in the STS theory, which plays a role equivalent to the energy eigenvalue equation in the usual QM. We have seen that, similar to the traditional QM, where time-dependent potentials yield time-dependent energy eigenstates, in the STS extension, position-dependent potentials produce position-dependent momentum eigenstates.

Then, we proposed an interpretation of the solutions of the SC Schr\"odinger equation inspired by the role of the Schr\"odinger equation in the usual QM: Given an ``initial'' condition $\phi(t|x_0)$, which represents the time probability amplitude of the particle arriving at $x_0$, the solution of the SC Schrödinger equation provides the time probability amplitude of the particle arriving at $x \lessgtr x_0$. In this scenario, whereas in the standard QM, to experimentally obtain $|\psi(x|t)|^2$, the detector must be spread in space and switched on at time $t$, in the STS extension, to experimentally obtain $|\phi(t|x)|^2$, the detector must be localized at position $x$ and switched on all the time. Thus, while the SC Schr\"odinger equation tells how probabilities change with the position of the detector, the Schr\"odinger equation provides how probabilities change with the time established for the detector to measure the particle.

To make this interpretation more transparent, we described the solution of the SC Schrödinger equation in terms of a space ``evolution'' operator, akin to the time evolution operator of standard QM. To do this, we defined a space-ordering operator (analogous to the time-ordering operator of the usual QM), which orders operations in terms of either increasing or decreasing values of $x$.

We also proposed an interpretation of the eigenstates of the STS extension. We focused on the differences between the eigenstates of the same observable in the STS extension and in standard QM. Specifically, we highlighted a crucial distinction: in standard QM, $|P\rangle\equiv |P\rangle|_t$ signifies the state of a particle with momentum $P$ (and indefinite position) at time $t$. Besides, $|P\rangle|_t$ maintains its mathematical form regardless of the applied potential, given by the Fourier transform of $|x\rangle$. On the other hand, in the STS extension, $|P_b(x)\rangle$ represents the state of a particle with momentum $P_b(x)$ at position $x$ --- meaning that the particle arrives at position $x$ with momentum $P_b(x)$ (at an indefinite arrival time). Importantly, unlike $|P\rangle|_t$, $|P_b(x)\rangle$ depends on the potential involved and must be determined through the eigenvalue equation~(\ref{SchroTP1}). When incorporating these interpretations into the state of the particle $|\phi(x)\rangle$, we verified that the momentum wave function in the STS extension, ${\tilde \phi}(b|x)$, becomes the probability amplitude of the particle arriving at position $x$ with momentum $P_b(x)$.


After considering these interpretations of the STS extension and examining some illustrative examples, we concluded that $|\psi(t)\rangle$ and $|\phi(x)\rangle$ provide complementary information about the same particle. In this manner, the quantum state of the particle at time $t$ is as fundamental as its quantum state at position $x$.


Finally, we solved the SC Schr\"odinger equation for an arbitrary $V=V(x)$, allowing us to predict traversal and tunneling times of a potential barrier.  From this investigation, we concluded that if we do not establish a connection between the momentum wave functions of the STS extension and the usual QM, the predictions of the STS extension can diverge from the Kijowski distribution. Further exploration of the physical implications of Eq.~(\ref{SolTX}) and the pursuit of an operational procedure involving detectors and clocks in standard quantum mechanics to obtain predictions for $|\phi(x)\rangle$ are left for future work. Additionally, the investigation of the three-dimensional version of the STS extension proposed in Ref.~\cite{Dias3} is deferred to future research


\section{\label{sec:level8}acknowledgements}
E.O.D. acknowledges financial support from CNPq (Conselho Nacional de Desenvolvimento
Cient\'ifico e Tecnol\'ogico) through Program No. 09/2020 (Grant No. 315759/2020-8). R.X. acknowledges financial support from the U.S. Department of Energy under contract number DE-SC0017647. R.E.A. acknowledges financial support from CNPq.

\end{document}